# Universality of BSW mechanism for spinning particles

Jie Jiang[a], Sijie Gao[b]

Department of Physics, Beijing Normal University, Beijing 100875, China

**Abstract** Bañados et al. (BSW) found that Kerr black holes can act as particle accelerators with collisions at arbitrarily high center-of-mass energies. Recently, collisions of particles with spin around some rotating black holes have been discussed. In this paper, we study the BSW mechanism for spinning particles by using a metric ansatz which describes a general rotating black hole. We notice that there are two inequivalent definitions of center-of-mass (CM) energy for spinning particles. We mainly discuss the CM energy defined in terms of the worldline of the particle. We show that there exists an energy-angular momentum relation $e = \Omega_h j$ that causes collisions with arbitrarily high energy near extremal black holes. We also provide a simple but rigorous proof that the BSW mechanism breaks down for nonextremal black holes. For the alternative definition of the CM energy, some authors find a new critical spin relation that also causes the divergence of the CM mass. However, by checking the timelike constraint, we show that particles with this critical spin cannot reach the horizon of the black hole. Further numerical calculation suggests that such particles cannot exist anywhere outside the horizon. Our results are universal, independent of the underlying theories of gravity.

## 1 Introduction

In 2009, Bañados, Silk and West [1] first showed that extremal Kerr black holes can act as accelerators of particles and arbitrarily high center-of-mass (CM) energies can be obtained if one of the particles satisfies $j = 2e$, where $j$ and $e$ are angular momentum and energy per unit mass. The BSW mechanism then was extended to a variety of black holes [2–14]. Most studies so far focus on test particles. A more realistic model is to treat the particle as an extended body with self-interaction and take the particle's spin into account. The orbit of a spinning particle deviates from a geodesic and can be described by the Mathisson–Papapetrou–Dixon (MPD) equation [15–18]. The BSW mechanism for spinning particles in Kerr, Kerr–Newman and Kerr–Sen spacetimes has been investigated [19–21].

[a] e-mail: jiejiang@mail.bnu.edu.cn
[b] e-mail: sijie@bnu.edu.cn

The aim of our paper is to explore this issue in a model-independent way. A general stationary axisymmetric black hole can be described by the metric ansatz

$$ds^2 = \alpha(r,\theta)^2 \left( -\Delta(r,\theta) dt^2 + \frac{dr^2}{\Delta(r,\theta)} \right)$$
$$+ \frac{d\theta^2}{\Delta_\theta(r,\theta)} + A(r,\theta)^2 \left( d\phi - \Omega(r,\theta) dt \right)^2, \quad (1)$$

which represents most stationary axisymmetric black holes in general relativity or other theories of gravity, such as: Kerr–AdS (dS) [22], warped AdS [23], Myers–Perry black hole [24], Kerr–Newman [20], Kerr–MOG [25,26], Kerr–Sen [27], Einstein–Maxwell-dilaton black holes [28], Horava–Lifshitz black holes [29], rotating charged cylindrical black holes [30], rotating ABG black holes [31], rotating black holes in a Randall–Sundrum brane [32], Kerr–Taub–NUT [33], Kerr–Bardeen black holes [34], rotating Hayward black holes [35,36], Bardeen black holes [37]. Zaslavskii [4] used a metric in coordinates $\{t, l, z, \phi\}$, which we believe is equivalent to Eq. (1), to obtain some general properties of the BSW mechanism for spinless particles.

Our analysis is based on the metric ansatz (1). We only impose a few natural conditions on the spacetime, like the existence of the equatorial plane located at $\theta = \pi/2$. We derive the general critical relation $j = \Omega_h e$, where $j$ and $e$ are the angular momentum and energy per unit mass of the particle, and $\Omega_h$ is the angular velocity of the horizon. This relation leads to the divergence of the CM energy around extremal black holes. Besides, we derive a lower bound on the energy per unit mass of the particle, which is 1/3 for spinless particles in Kerr spacetime.

It is worth noting that there are two inequivalent definitions of the CM energy in the literature. Usually, particle's momen-

Springer

tum is parallel to its worldline. But this is no longer true in the presence of spin. So the CM energy can be defined in terms of the worldline (see [19]) or the momentum (see [21]). We prefer the worldline version of the CM energy because if it diverges, it means that the relative speed of the two particles approaches the speed of light. In contrast, if the momentum version of the CM energy diverges, its physical implication is not very clear. Therefore, our analysis is mainly based on the worldline version, which we refer to as Definition 1. At the end of the paper, we calculate the BSW effect based on the momentum version (Definition 2). Interestingly, the two definitions give the same critical angular momentum-energy relation that causes the CM energy to diverge at the horizon. Definition 2 also leads to a critical spin, causing divergence of the CM energy in Kerr–Sen black holes [21]. However, our analysis shows that such divergence is unlikely to happen because it violates the timelike constraint of the particle.

The structure of the paper is as follows. In Sect. 2, we derive the motion of spinning particles in the equatorial plane of the black hole by employing the Mathisson–Papapetrou–Dixon equation. In Sect. 3, we derive the divergence condition of the CM energy associated with Definition 1 for extremal black holes and show that the BSW mechanism breaks down for nonextremal black holes. In Sect. 3, we investigate the divergence conditions associated with Definition 2. Concluding remarks are given in Sect. 5.

## 2 Motions of spinning particles

Let $F(r, \theta)$ represent one of the functions in Eq. (1). We assume that the spacetime possesses the parity reflection symmetric $\theta \to \pi - \theta$. This implies that in the equatorial place $\theta = \pi/2$,

$$\left.\frac{\partial F}{\partial \theta}\right|_{\theta=\pi/2} = 0, \tag{2}$$

In general, the black hole horizon is located at $r = r_h$ which satisfies $\Delta|_{r_h} = 0$. The angular velocity of the black at the horizon is

$$\Omega_h = \Omega(r_h). \tag{3}$$

The spacetime is stationary and axisymmetric with Killing vector fields

$$\xi^a = \left(\frac{\partial}{\partial t}\right)^a, \quad \phi^a = \left(\frac{\partial}{\partial \phi}\right)^a. \tag{4}$$

We shall investigate the behaviors of spinning particles moving outside the black hole. By integrating the stress-energy tensor $T^{ab}$ over the particle, one can define the momentum $P^a$ and the spin tensor $S^{ab}$ [15]. As shown by Beiglböck [41], there exists a unique timelike world line $x^\mu(\tau)$ which is called the center of mass of the particle. Then the motion of the particle can be described by the well-known Mathisson–Papapetrou–Dixon (MPD) equation [15,19]

$$\frac{DP^a}{D\tau} = -\frac{1}{2}R^a{}_{bcd}v^b S^{cd}, \tag{5}$$

$$\frac{DS^{ab}}{D\tau} = P^a v^b - P^b v^a, \tag{6}$$

where

$$v^a = \left(\frac{\partial}{\partial \tau}\right)^a \tag{7}$$

is the tangent vector to the world line and $\frac{D}{D\tau}$ is the covariant derivative along the world line. Define

$$u^a = \frac{P^a}{m} \tag{8}$$

as the dynamical velocity, where the mass $m$ is defined by $m^2 = -P^a P_a$, which is a constant along the orbit [15]. Using the normalization condition $u^a v_a = -1$, one can obtain the relation between $u^a$ and $v^a$ [42]:

$$v^a - u^a = \frac{S^{ab}R_{bcde}u^c S^{de}}{2\left(m^2 + \frac{1}{4}R_{bcde}S^{bc}S^{de}\right)}, \tag{9}$$

which implies that the momentum $P^a$ is not tangent to the world line of the spinning particle in curved spacetimes. Moreover, the two Killing vector fields in Eq. (4) correspond to the conserved energy and angular momentum per unit mass, which can be written as

$$e = \frac{E}{m} = -u^a \xi_a + \frac{1}{2m} S^{ab} \nabla_b \xi_a,$$

$$j = \frac{J}{m} = u^a \phi_a - \frac{1}{2m} S^{ab} \nabla_b \phi_a. \tag{10}$$

In order to calculate these quantities, we rewrite the metric in the tetrad form

$$g_{ab} = \eta_{\mu\nu} e_a^{(\mu)} e_b^{(\nu)}, \tag{11}$$

where

$$e_a^{(0)} = \alpha\sqrt{\Delta}(dt)_a,$$
$$e_a^{(1)} = \frac{\alpha}{\sqrt{\Delta}}(dr)_a,$$
$$e_a^{(2)} = \frac{1}{\sqrt{\Delta_\theta}}(d\theta)_a,$$
$$e_a^{(3)} = A\left((d\phi)_a - \Omega(dt)_a\right). \tag{12}$$

We consider spinning particles moving in the equatorial plane ($\theta = \frac{\pi}{2}$), which means the tetrad component $v^{(2)} = 0$. We first introduce a spin vector $s^{(a)}$ such that

$$S^{(a)(b)} = m\varepsilon^{(a)(b)}{}_{(c)(d)} u^{(c)} s^{(d)}, \tag{13}$$

where $\varepsilon_{(a)(a)(c)(d)}$ is the completely antisymmetric tensor with the component $\varepsilon_{(1)(2)(3)(4)} = 1$. From Eq. (9), we find $u^{(2)} = 0$ and the only nonzero component of $s^{(a)}$ is [42]

$$s^{(2)} = -s, \tag{14}$$

where $s$ indicates both the magnitude and the direction of the particle's spin, which is parallel to the black hole's spin for $s > 0$ and antiparallel for $s < 0$. Then we can show that the nonvanishing components of the antisymmetric spin tensor $S^{ab}$ are

$$S^{(0)(1)} = -msu^{(3)},$$
$$S^{(0)(3)} = msu^{(1)},$$
$$S^{(1)(3)} = msu^{(0)}. \tag{15}$$

Substituting Eq. (15) into Eq. (10), we obtain

$$e = \left(\frac{s\Omega A'}{\alpha} + \frac{As\Omega'}{2\alpha} + \alpha\right)\sqrt{\Delta}u^{(0)}$$
$$+ \left(A\Omega + \frac{\Delta s\alpha'}{\alpha} + \frac{s\Delta'}{2} - \frac{A^2 s\Omega\Omega'}{2\alpha^2}\right)u^{(3)}, \tag{16}$$

$$j = \frac{\sqrt{\Delta}sA'}{\alpha}u^{(0)}$$
$$+ \left(\frac{A^2 s\Omega'}{2\Omega^2(\alpha\Omega - 1)} - \frac{A}{\Omega} + \frac{1}{2}\alpha s\Delta'\right)u^{(3)}, \tag{17}$$

where the prime represents the derivative with respect to $r$. Here we have used Eq. (2), i.e., the partial derivative with respect to $\theta$ vanishes on the equatorial plane. In the tetrad frame, the normalization condition $u^a u_a = -1$ is simply

$$-\left(u^{(0)}\right)^2 + \left(u^{(1)}\right)^2 + \left(u^{(3)}\right)^2 = -1. \tag{18}$$

Solving Eqs. (16)–(18), we find

$$u^{(0)} = \frac{\Lambda}{\sqrt{\Delta}X}, \tag{19}$$

$$u^{(1)} = \frac{\sqrt{O}}{\sqrt{\Delta}X}, \tag{20}$$

$$u^{(3)} = \frac{Y}{X}, \tag{21}$$

where

$$O = \Lambda^2 - \Delta\left(X^2 + Y^2\right), \tag{22}$$

and

$$\Lambda = 2js\alpha^2(2\Delta\alpha' + \alpha\Delta')$$
$$+ 2A\alpha(e - j\Omega)(As\Omega' - 2\alpha^2), \tag{23}$$
$$X = A^3 s^2 \left(\Omega'\right)^2 + 2\alpha^2 s^2 A'\Delta' + 4\alpha\Delta s^2 \alpha' A' - 4\alpha^4 A, \tag{24}$$
$$Y = -2\alpha^2\left[Ajs\Omega' + 2\alpha^2 j - 2sA'(e - j\Omega)\right]. \tag{25}$$

By calculating Eq. (9), the relation between the normalized momentum vector $u^a$ and the 4-velocity $v^a$ can be expressed as

$$v^{(0)} \propto (X + \Delta\xi)u^{(0)} + \sqrt{\Delta}\zeta u^{(3)}, \tag{26}$$
$$v^{(1)} \propto Xu^{(1)}, \tag{27}$$
$$v^{(3)} \propto \omega u^{(3)} - \sqrt{\Delta}\zeta u^{(0)}, \tag{28}$$

where

$$\xi = 4\alpha s^2\left(\alpha A'' - 2\alpha' A'\right), \tag{29}$$
$$\zeta = -2As^2\left[\Omega'\left(3\alpha A' - 2A\alpha'\right) + \alpha A\Omega''\right], \tag{30}$$
$$\omega = 2\alpha^2 s^2 A\Delta'' - 4\alpha^4 A - 3A^3 s^2\left(\Omega'\right)^2 + 4A\alpha s^2\alpha'\Delta'$$
$$+ \Delta A\left(4\alpha s^2\alpha'' - 4s^2 A\left(\alpha'\right)^2\right). \tag{31}$$

For spinning particles, $v^a v_a$ is not a conserved quantity. This will lead to possible superluminal motions and causality problems. Therefore, we need to impose the timelike condition

$$v^a v_a < 0, \tag{32}$$

i.e.,

$$X^4 + X^2 Y^2 + 2\Lambda X(\Lambda\xi + \zeta Y)$$
$$- (\zeta\Lambda - \omega Y)^2 + \Delta(\Lambda\xi + \zeta Y)^2 > 0. \tag{33}$$

Finally Eq. (19) implies that a physically allowed trajectory should satisfy

$$O \geq 0. \tag{34}$$

## 3 Collision of two spinning particles

As mentioned in the introduction, there are two inequivalent definitions of the CM energy. We shall use the definition in terms of the worldline (Definition 1) [19] in this section. The alternative definition will be discussed in the next section.

### 3.1 Center-of-mass energy

Now we consider two spinning particles with the same mass $m$ colliding into each other outside the black hole. The center-of-mass energy is defined by [1]

$$E_{c.m.} = \sqrt{2}m\sqrt{1 - g_{ab}n_1^a n_2^b}, \tag{35}$$

where $n^a = \lambda v^a$ is the normalization of $v^a$, satisfying $n_a n^a = -1$, and the subscripts 1 and 2 label the two particles. For simplicity, we can define the effective CM energy [3]

$$E_{\text{eff}} \equiv -g_{ab}n_1^a n_2^b = -\lambda_1\lambda_2 g_{ab}v_1^a v_2^b. \tag{36}$$

Let

$$V_{ij} = -g_{ab}v_i^a v_j^b, \tag{37}$$

where $i, j = 1, 2$. Then it is easy to show

$$E_{\text{eff}} = \frac{V_{12}}{\sqrt{V_{11} V_{22}}}. \tag{38}$$

Substituting Eqs. (19) and (26) into Eq. (37), one finds

$$V_{12} = \frac{\Lambda_1 \Lambda_2 - \sqrt{O_1 O_2}}{\Delta} + \frac{B_{12} + \Delta C_{12}}{X_1 X_2}, \tag{39}$$

$$V_{ii} = X_i^2 + Y_i^2 + \frac{B_{ii} + \Delta C_{ii}}{X_i^2}, \tag{40}$$

where

$$B_{ij} = \zeta_j \Lambda_i X_i Y_j + \zeta_i \Lambda_j X_j Y_i - \zeta_i \zeta_j \Lambda_i \Lambda_j + \Lambda_i \Lambda_j \xi_j X_i$$
$$+ \Lambda_i \Lambda_j \xi_i X_j + \zeta_j \Lambda_j \omega_i Y_i + \zeta_i \Lambda_i \omega_j Y_j - \omega_i \omega_j Y_i Y_j, \tag{41}$$

$$C_{ij} = (Y_i \zeta_i + \Lambda_i \xi_i)(Y_j \zeta_j + \Lambda_j \xi_j). \tag{42}$$

Our purpose is to investigate whether arbitrarily high CM energies can be obtained near the black hole. By substituting Eqs. (39) and (40) into Eq. (38), we see that a divergent CM energy is attainable only on the horizon $\Delta = 0$. Therefore, we need to study the behavior of $E_{\text{eff}}$ near the horizon. Denote the first numerator in Eq. (39) by $K$, i.e.,

$$K = \Lambda_1 \Lambda_2 - \sqrt{O_1 O_2}. \tag{43}$$

By expanding $K$ near the horizon, we find that the leading term is given by

$$K \approx \frac{\Delta}{2} \left( \frac{(X_1^2 + Y_1^2) \Lambda_2}{\Lambda_1} + \frac{(X_2^2 + Y_2^2) \Lambda_1}{\Lambda_2} \right). \tag{44}$$

Substituting Eq. (44) into Eq. (39), we get

$$V_{12}|_h = \frac{1}{2} \left( \frac{(X_1^2 + Y_1^2) \Lambda_2}{\Lambda_1} + \frac{(X_2^2 + Y_2^2) \Lambda_1}{\Lambda_2} \right) + \frac{B_{12}}{X_1 X_2}, \tag{45}$$

and

$$V_{ii}|_h = X_i^2 + Y_i^2 + \frac{B_{ii}}{X_i^2}, \tag{46}$$

where $h$ denotes the horizon. It is easy to see that the necessary condition for $E_{\text{eff}}$ to blow up is that one of the particles satisfies $\Lambda = 0$ at the horizon. Note that near the horizon, $\Delta \sim (r - r_h)^2$ for the extremal case, and $\Delta \sim (r - r_h)$ for the non-extremal case. It means that $\Delta' = 0$ on the extremal horizon and $\Delta' \neq 0$ for the non-extremal horizon. This difference is crucial on the BSW mechanism. Therefore, we need to discuss them respectively.

### 3.2 Extremal case

In the extremal case, using $\Delta = \Delta' = 0$ on the horizon, Eq. (23) becomes

$$\Lambda = 2A\alpha(e - j\Omega_h)\left[A\Omega' s - 2\alpha^2\right], \tag{47}$$

$$X = A^3 \Omega'^2 s^2 - 4A\alpha^4. \tag{48}$$

Denote

$$s_c = \frac{2\alpha^2}{A\Omega'}\bigg|_h. \tag{49}$$

Then

$$\Lambda = 2A^2 \Omega' \alpha (e - j\Omega_h)(s - s_c), \tag{50}$$

$$X = A^3 \Omega'^2 (s^2 - s_c^2), \tag{51}$$

and

$$V_{12} = F_1 \frac{(e_2 - j_2 \Omega_h)(s_2 - s_c)}{(e_1 - j_1 \Omega_h)(s_1 - s_c)}$$
$$+ F_2 \frac{(e_1 - j_1 \Omega_h)(s_1 - s_c)}{(e_2 - j_2 \Omega_h)(s_2 - s_c)} + \frac{H_{12}}{(s_1^2 - s_c^2)(s_2^2 - s_c^2)}, \tag{52}$$

$$V_{ii} = 2F_i + \frac{H_{ii}}{(s_i^2 - s_c^2)^2}, \tag{53}$$

where

$$F_i = \frac{1}{2}(X_i^2 + Y_i^2), \quad H_{ij} = \frac{B_{ij}}{A^6 \Omega'^4}. \tag{54}$$

Without loss of generality, we assume that $s_i^2 < s_c^2$. Substitution of Eqs. (52) and (53) into Eq. (38) yields

$$E_{\text{eff}} = \frac{(s_1^2 - s_c^2)(s_2^2 - s_c^2)}{\sqrt{[H_{11} + 2F_1(s_1^2 - s_c^2)^2][H_{22} + 2F_2(s_2^2 - s_c^2)^2]}}$$
$$\times \left[ \frac{H_{12}}{(s_1^2 - s_c^2)(s_2^2 - s_c^2)} + \frac{F_1(s_2 - s_c)(e_2 - \Omega_h j_2)}{(s_1 - s_c)(e_1 - \Omega_h j_1)} \right.$$
$$\left. + \frac{F_2(s_1 - s_c)(e_1 - \Omega_h j_1)}{(s_2 - s_c)(e_2 - \Omega_h j_2)} \right], \tag{55}$$

which shows clearly that $E_{\text{eff}}$ is divergent if and only if one of the particle satisfies the relation

$$e_i = \Omega_h j_i. \tag{56}$$

This is the general critical condition which has been confirmed by many black hole solutions.

Finally, we need to ensure that such a particle can actually reach the horizon. Note that when Eq. (56) is satisfied, we have $\Lambda|_h = 0$. Then, condition (33) becomes

$$X^4 + X^2 Y^2 - Y^2 \omega^2 > 0 \quad \text{on } r = r_h, \tag{57}$$

i.e.

$$\left(A^3 s^2 (\Omega')^2 - 4\alpha^4 A\right)^4 + 4\alpha^4 A^2 j^2$$
$$\times \left(2\alpha^2 + As\Omega'\right)^2 \left(A^2 s^2 (\Omega')^2 - 4\alpha^4\right)^2$$
$$-4\alpha^4 j^2 \left(2\alpha^2 + As\Omega'\right)^2$$
$$\times \left(3A^3 s^2 (\Omega')^2 - 2\alpha^2 As^2 A'' + 4\alpha^4 A\right)^2 > 0. \quad (58)$$

Since $O'|_h = 0$ and

$$O''|_h = 2\Lambda'^2 - \Delta''(X^2 + Y^2), \quad (59)$$

condition.(34) can be written as

$$2\Lambda'^2 - \Delta''(X^2 + Y^2) > 0, \quad (60)$$

i.e.

$$8\alpha^2 \left(\alpha^2 js\Delta'' - Aj\Omega'\left(As\Omega' - 2\alpha^2\right)\right)^2 - \Delta''$$
$$\times \left(\left(A^3 s^2 \Omega^2 - 4\alpha^4 A\right)^2 + 4\alpha^4 \left(Ajs\Omega + 2\alpha^2 j\right)^2\right) > 0 \quad (61)$$

on the horizon $r = r_h$.

In particular, for the spinless case $s = 0$, one can show that Eq. (58) is always satisfied and Eq. (61) reduces to

$$e^2 > \frac{\Omega^2 \alpha^2 A^2 \Delta''}{2A^2 \Omega'^2 - \alpha^2 \Delta''}\Big|_h. \quad (62)$$

Note that the right-hand side of the inequality is a constant depending only on the parameters of the black hole. So in addition to the critical relation, the energy of the particle should satisfy this inequality as well. In Kerr–Newman spacetimes, Eq. (62) becomes

$$e^2 > \frac{M^2 - Q^2}{3M^2 - 4Q^2}, \quad (63)$$

where we have used the extremal conditions $r_h = M, a = \sqrt{M^2 - Q^2}$. For Kerr black holes $Q = 0$, Eq. (63) further reduces to $e^2 > \frac{1}{3}$. This lower bound imposed on the energy of the particle per unit mass is a necessary condition for the BSW mechanism, which has been ignored in the literature.

### 3.3 Nonextremal case

We may repeat the above computation for nonextremal cases. The major difference is that $\Delta' \neq 0$ for nonextremal black holes. By similar calculation, we obtain now

$$E_{\text{eff}} = \frac{2B_{12}\Lambda_1\Lambda_2 + X_1^3 X_2 \Lambda_2^2 + X_1 X_2^3 \Lambda_1^2 + X_1 X_2 Y_2^2 \Lambda_1^2 + X_1 X_2 Y_1^2 \Lambda_2^2}{2\sqrt{B_{11} + X_1^4 + X_1^2 Y_1^2}\sqrt{B_{22} + X_2^4 + X_2^2 Y_2^2}\Lambda_1\Lambda_2}. \quad (64)$$

Eq. (64) shows that the necessary condition for the CM energy to blow up is

$$\Lambda|_h = 0. \quad (65)$$

However, we need to check Eq. (34) to see whether the particle can reach the horizon under this condition. From Eq. (22), we see immediately

$$O|_h = \Lambda^2 - (X^2 + Y^2)\Delta = 0. \quad (66)$$

Furthermore,

$$O'|_h = 2\Lambda\Lambda' - (X^2 + Y^2)'\Delta$$
$$- (X^2 + Y^2)\Delta' = -(X^2 + Y^2)\Delta'. \quad (67)$$

Since

$$\Delta'|_h > 0, \quad (68)$$

we have

$$O \approx O'|_h (r - r_h) < 0, \quad (69)$$

near the horizon, which means that the particle cannot reach the horizon with the parameters satisfying the condition (65). So the divergent energy can not be obtained mechanism fails for non-extremal black holes when the particle satisfies the critical relation (64). Although this result has been known for many specific black holes, we provide the general proof for the first time and only use the simple fact that $\Delta$ is positive outside the black hole.

However, as shown in Refs. [4,38–40], For non-extremal black holes, the arbitrary high energy can also be obtained from the collision in the near-horizon region when one of the particles possesses the near-critical angular momentum. Here, we shall extend this argument to our general framework. Let $r = r_c$ be the point where the collision occurs. Assume the particle satisfies the so-called near-critical relation

$$\Lambda(r_c) = \chi\sqrt{\Delta(r_c)}, \quad (70)$$

where $\chi$ is a finite coefficient satisfying

$$(\chi^2 - X^2 - Y^2)\Big|_{r_c} > 0, \quad (71)$$

which ensures that $O(r_c) > 0$.

When particle 1 is near-critical, i.e., satisfying Eq. (70), and particle 2 is usual, by using Eqs. (38) and (39), we can obtain the effective energy at the collision point

$$E_{\text{eff}} \approx \frac{\Lambda_2|_h}{\sqrt{\Delta(r_c)}} \left[\chi_1 - \sqrt{\chi_1^2 - X_1^2 - Y_1^2}\right]\Big|_h. \quad (72)$$

We see that $E_{\text{eff}}$ is divergent when $\Delta(r_c) \to 0$.

Finally, in order to avoid superluminal motion of the near-critical particle, we should consider the timelike condition (33), which reduces to

$$X^4 + X^2Y^2 - \omega^2 Y^2 > 0. \tag{73}$$

at near the horizon. In the Schwarzchild spacetime, the critical relation (65) becomes

$$e = \frac{Msj}{r_h^3} = \frac{sj}{8M^2}, \tag{74}$$

Substitution Eq. (74) into Eq. (73) yields

$$e < \frac{16M^2 - 2s^2}{8M\sqrt{3(16M^2 + s^2)}} < \frac{1}{2\sqrt{3}} < 1. \tag{75}$$

These results are actually in agreement with those found in Ref. [38]. Obviously, a particle released from infinity must satisfy $e \geq 1$. Thus, Eq. (75) implies that the collision cannot be realized for particles falling from infinity. This differs from the extremal case. As suggested by Zaslavskii [4,38], the near-critical particle can be produced through multiple collisions, instead of direct collisions.

## 4 Collision of two spinning particles—alternative definition of CM energy

In this section, we turn to an alternative definition (Definition 2) of the CM energy, which is [20,39]

$$E'_{c.m.} = \sqrt{2}m\sqrt{1 - g_{ab}u_1^a u_2^b}, \tag{76}$$

where $u^a$ is given by Eq. (8), i.e., proportional to the momentum $P^a$. The corresponding effective CM energy is just

$$E'_{\text{eff}} = -g_{ab}u_1^a u_2^b. \tag{77}$$

Substituting Eq. (19) into Eq. (77), one obtains

$$E'_{\text{eff}} = \frac{\Lambda_1 \Lambda_2 - \sqrt{O_1 O_2}}{\Delta} - \frac{Y_1 Y_2}{X_1 X_2}. \tag{78}$$

We see that a divergent CM energy is possible only when $X_i = 0$ or $\Delta = 0$.

First, the effective energy at the horizon $\Delta = 0$ takes the form

$$E'_{\text{eff}} = \frac{1}{2}\left(\frac{(X_1^2 + Y_1^2)\Lambda_2}{\Lambda_1} + \frac{(X_2^2 + Y_2^2)\Lambda_1}{\Lambda_2}\right) + \frac{Y_1 Y_2}{X_1 X_2}. \tag{79}$$

In the extremal case $\Delta' = 0$, the divergence condition $\Lambda_i = 0$ yields

$$e_i = \Omega_h j_i, \tag{80}$$

which is the same as that we obtained in the last section. However, Eq. (78) indicates that when $X_i = 0$, $E'_{\text{eff}}$ also diverges. This gives another critical condition

$$s_i = -s_c. \tag{81}$$

This critical spin relation does not apply to the CM energy defined in the last section. As discussed in [21], this spin relation can cause infinite CM energy near Kerr–Sen black holes since the critical spin $s_c$ can be arbitrarily small. However, by substituting Eq. (81) and calculate $v^a v_a$ at the horizon, we find

$$v^a v_a \propto (\zeta \Lambda - \omega Y)^2 > 0, \tag{82}$$

which violates the timelike condition (33). Therefore, particles with the critical spin cannot reach the horizon and the divergence of the CM energy cannot be realized.

For the nonextremal cases, in the last section, we have proven that for the exactly critical relation $\Lambda = 0$, the particle cannot reach the horizon. One can check that the divergence condition $X = 0$ also violates the constraint (33) at the horizon, i.e., the divergent CM energy cannot be obtained for the nonextremal black hole through the collision on the horizon when the particle satisfied the critical relations. However, by the argument similar to Sect. 3.3, one can show that arbitrarily high CM energies can be obtained when one of the particle satisfies the near-critical relation (70) in nonextremal cases.

Since the ultraenergetic collision near the horizon has been ruled out, we now turn to the collision that occurs away from the horizon. By imposing the critical condition $X = 0$, we find that the timelike condition Eq. (34) becomes

$$\mathcal{T} = \Delta(\Lambda \xi + \zeta Y)^2 - (\zeta \Lambda - \omega Y)^2 > 0. \tag{83}$$

For Kerr–Newmann spacetime, we have $\xi = \zeta = 0$. So Eq. (83) cannot be satisfied anywhere outside the horizon.

An interesting example is the Kerr–Sen black hole, for which the corresponding parameters in the line element (1) can be written as

$$\alpha^2 = \frac{r(\mu\cosh(2\eta) - \mu + r)}{\sqrt{r\left(4\mu\sinh^2(\eta)\left(a^2 + r^2 + \mu r \sinh^2(\eta)\right) + a^2(2\mu + r) + r^3\right)}}, \tag{84}$$

$$\Delta = \frac{a^2 + r(r - 2\mu)}{\sqrt{\left(a^2 + r^2 + 2\mu r \sinh^2(\eta)\right)^2 - a^2\left(a^2 + r(r - 2\mu)\right)}}, \tag{85}$$

$$A^2 = \frac{a^2(2\mu\cosh(2\eta) + r)}{\mu\cosh(2\eta) - \mu + r} + \mu r\cosh(2\eta) + r(r - \mu), \tag{86}$$

$$\Omega = \frac{2a\mu\cosh^2(\eta)}{4\mu\sinh^2(\eta)\left(a^2 + r^2 + \mu r \sinh^2(\eta)\right) + a^2(2\mu + r) + r^3}, \tag{87}$$

where the parameters $\mu$, $\eta$, and $a$ are associated with the physical mass $M$, charge $Q$, and angular momentum $J$. Substituting Eqs. (85)–(87) into Eq. (83), together with the divergence condition $X = 0$, one can obtain the function $\mathcal{T}(r)$. The expression of $\mathcal{T}(r)$ is rather complicated and it is unlikely to make a complete analysis. However, numerical calculation suggests (see Fig. 1) that $\mathcal{T} - r$ is always negative outside the horizon, which means that, for the Kerr–Sen black hole, the critical spin relation (81) cannot lead to the divergence of the CM energy. This result is different from that in [21].

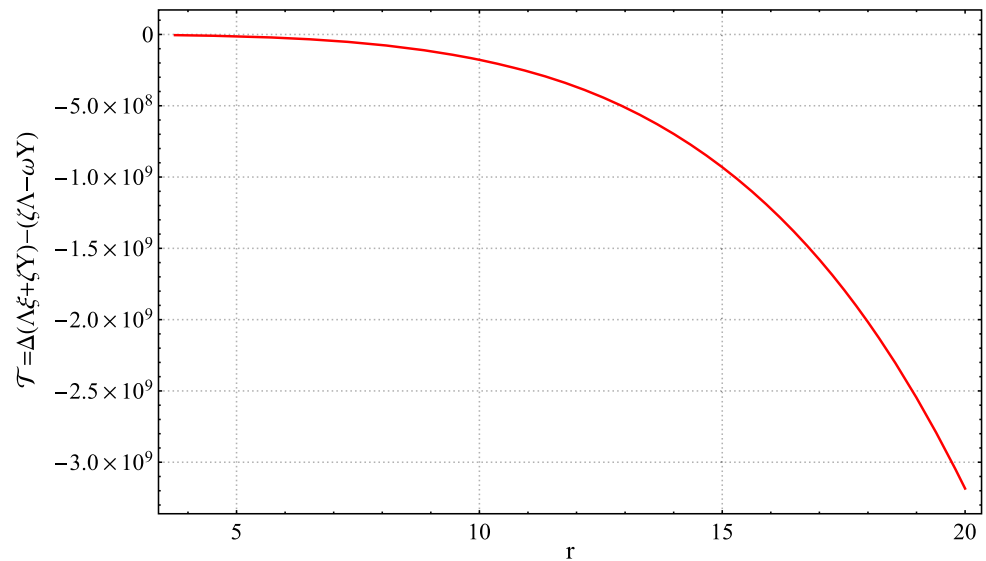

**Fig. 1** The plot of $\mathcal{T} - r$ out of the horizon, where we set $j = 2, e = 1, \mu = 2, a = 1, \eta = 1$

## 5 Concluding remarks

In summary, we have found some universal features of the BSW mechanism for spinning particles colliding near rotating black holes. We noticed that two different definitions of the CM energy have been used in the literature. For Definition 1, i.e., the definition in terms of the world line, we have shown that arbitrarily high CM energies can be obtained near the horizon of an extremal black hole when the critical relation $e = \Omega_h j$ is satisfied. Although the definitions of the conserved angular momentum and energy have been modified due to particle's spin, the critical relation is the same as spinless particles. In addition, we found another constraint on the particle's energy per unit mass. We also showed that collisions around a nonextremal black hole cannot create divergent CM energies when the particle satisfies the critical relation, but when one of the particles satisfies the near-critical relation (70), divergent CM energies can be obtained in the scenario of multiple scattering instead of direct collision.

For Definition 2 of the CM energy, we showed that the same critical relation $e = \Omega_h j$ gives the divergence of the CM energy near the extremal horizon. Some authors claimed that the critical spin relation $s = -s_c$ can also cause ultraenergetic collisions around Kerr–Sen black holes. However, by checking the timelike condition, we demonstrated that such collisions are unlikely to happen.

Our analyses are mainly based on a general metric ansatz describing rotating black holes. No field equations have been employed in our derivation. So our results are theory-independent and can apply to a wide class of rotating black holes.

**Acknowledgements** This research was supported by NSFC Grant nos. 11775022 and 11375026.

**Data Availability Statement** This manuscript has no associated data or the data will not be deposited. [Authors' comment: The paper has no associated data as it is a theoretical paper.]